\documentclass[]{aastex701}
\usepackage{soul}
\usepackage{graphicx}
\usepackage{hyperref}
\usepackage[table,xcdraw]{xcolor}

\received{August 23, 2025}
\revised{October 8, 2025}
\accepted{\today}

\submitjournal{ApJ}

\graphicspath{{./}{figures/}}

\begin{document}

\title{Statistical Overview of Long-Lived Active Regions Observed Across Multiple Carrington Rotations}

%% To identify a corresponding author, use the \correspondingauthor command.
%% The command appends "Corresponding Author: " to the argument it appears at
%% the bottom left of the first page like the output from \email. 

\author[0000-0002-8767-7182]{Emily~I.~Mason}
\affiliation{Predictive Science Inc., 9990 Mesa Rim Rd., Ste. 170, San Diego, CA 92121, USA} 
\email[show]{emason@predsci.com}
\correspondingauthor{Emily I. Mason}

\author[0000-0003-3740-9240]{Kara L. Kniezewski}
\affiliation{Air Force Institute of Technology, 2950 Hobson Way, Wright-Patterson AFB, OH 45433, USA}
\email[show]{kara.kniezewski@us.af.mil}

\begin{abstract}

The study of solar active regions (ARs) is of central importance to a range of fundamental science, as well as the practical applications of space weather. Active region emergence and life cycles are two areas of particular interest, yet the lack of consistent full-Sun observations has made long-term studies of active regions difficult. Here, we present results from a study to identify and characterize long-lived active regions (LLARs), defined as those which were observed during at least two consecutive Carrington rotations and which did not undergo significant successive flux emergence once the decay phase began. Such active regions accounted for 13\% of all NOAA-identified ARs between 2011 and 2019, and their distribution closely follows the annual sunspot number. This implies that LLARs are produced by the same basic driving processes as regular ARs. LLAR areas tend to be significantly larger and contain more magnetic flux compared to other ARs, but the two categories have similar magnetic complexity distributions. The most striking result, however, is that LLARs are 3-6 times more likely than other ARs to be the source of a solar flare of GOES class C or greater. This highlights the importance of studying what makes a LLAR and how to identify them at emergence with a view towards improved space weather forecasting. The further implications of these findings for AR heating spatial and temporal patterns will be explored in an upcoming study.

% note to self: check rates of anti-Hale among LLARs vs. all ARs. Also flare activity.

\end{abstract}


\section{Introduction}\label{sec:intro}

Solar active regions (ARs) -- vast concentrations of magnetic field that emerge through the photosphere and then gradually dissipate over days to months -- are structures of enduring research interest within the solar physics community. They are the source location for solar flares \citep[][and references therein]{Benz2017} and many coronal mass ejections \citep[][and references therein]{Chen2011,Webb2012}, and other energetic phenomena which can have myriad effects on human life and technologies \citep[e.g.,][]{Pulkkinen2007,Reames2017,Parker2024}.

ARs have enormous ranges in characteristics. On one end of the spectrum, an AR can appear in a magnetogram as a small, simple pair of flux concentrations known as a bipole which emerges and decays within a few hours with no eruptive activity whatsoever. On the other end of the spectrum are quadrupolar or octopolar ARs with multiple neutral lines, which can experience large axial rotations or high rates of shearing. These ARs can take weeks or months to decay, and often host many flares. In between lie most ARs -- as well as outliers like NOAA 12192, which was very large and flare-productive but not particularly magnetically complex, remaining essentially one large bipole for most of its evolution. This large variety makes it both necessary and challenging to identify what drives ARs' creation, evolution, and decay.

The AR identifying scheme that is used most widely across the field of heliophysics is the NOAA active region number. Starting from 1972, sunspot groups which have been observed by two official observatories have been given a sequential 5-digit designation (though for various technical reasons, very often the first digit is dropped). There are several complications that this method engenders:
\begin{itemize}
    \item This system does not actually directly identify ARs, but rather sunspots. While the two phenomena are closely related, sunspots often vanish from visible wavelengths well before the magnetic field has decayed fully, such that it is indistinguishable from the background quiet Sun fields. Other systems that are magnetogram-based, such as the Space-Weather HMI Active Region Patches \citep[SHARPs;][]{Bobra2014}, identify ARs by their fields instead and therefore identify many more regions -- either those which are never strong enough to generate sunspots, or that have decayed sufficiently that their sunspots are no longer detectable. This is preferable for long-term studies on the full life cycle of ARs.
    \item There can be ambiguity around the designations when dropping the first digit, given that numbering ``reset'' in 2002 when the 10,000th AR was identified.
    \item If an AR lasts more than one rotation, it receives a new, uncorrelated numerical designation when it becomes Earth-facing again. Often, the daily NOAA Solar Region Summary will mention that an already-developed AR coming into view is likely a previously identified AR and may even list its previous designation. This is sporadic at best, however, and difficult to quickly parse into a coherent list of multi-rotation ARs. It also cannot account for other ARs in the same neighborhood which may have emerged during the intervening half of a solar rotation.
\end{itemize}

This last issue is the one that this project addresses, by compiling a list of ARs which survived several solar rotations and received multiple NOAA AR designations. Since the goal is to enable more statistical studies of ARs' lifetimes and decay phases, we added an extra constraint when creating the list and eliminated those long-lived ARs (LLARs) which experienced a significant secondary phase of flux emergence after their original formation. Section \ref{sec:methods} addresses how we identified possible LLARs and how the final list was generated. The results and statistics obtained from a preliminary investigation of the LLARs' characteristics are presented in Section \ref{sec:results}, and a discussion of the results and future plans can be found in Section \ref{sec:disc}.

\section{Methods}\label{sec:methods}

\subsection{Data}\label{sec:data}

For this project, we used a list of all NOAA-identified ARs between 2011 January 1 and 2019 December 31, compiled from the Heliophysics Event Knowledgebase (HEK, \cite{Hurlburt2012}) using {\tt SunPy} \citep{Barnes2020}. For each AR initially identified, its anticipated re-emergence time and location at the east limb was calculated with the differential rotation tools in {\tt SunPy}. A list of pairs of ARs was generated in this way for all the ARs found in a 200 x 200 arcsec box centered on the original AR's Carrington coordinates, within a time window of $\pm$3 days from the expected reappearance date (if multiple ARs were in the 200 x 200 arsec box, each candidate pair was listed separately for evaluation). The ARs identified on the subsequent rotation were the candidate matches.

\begin{figure}
    \centering
    \includegraphics[width=\linewidth]{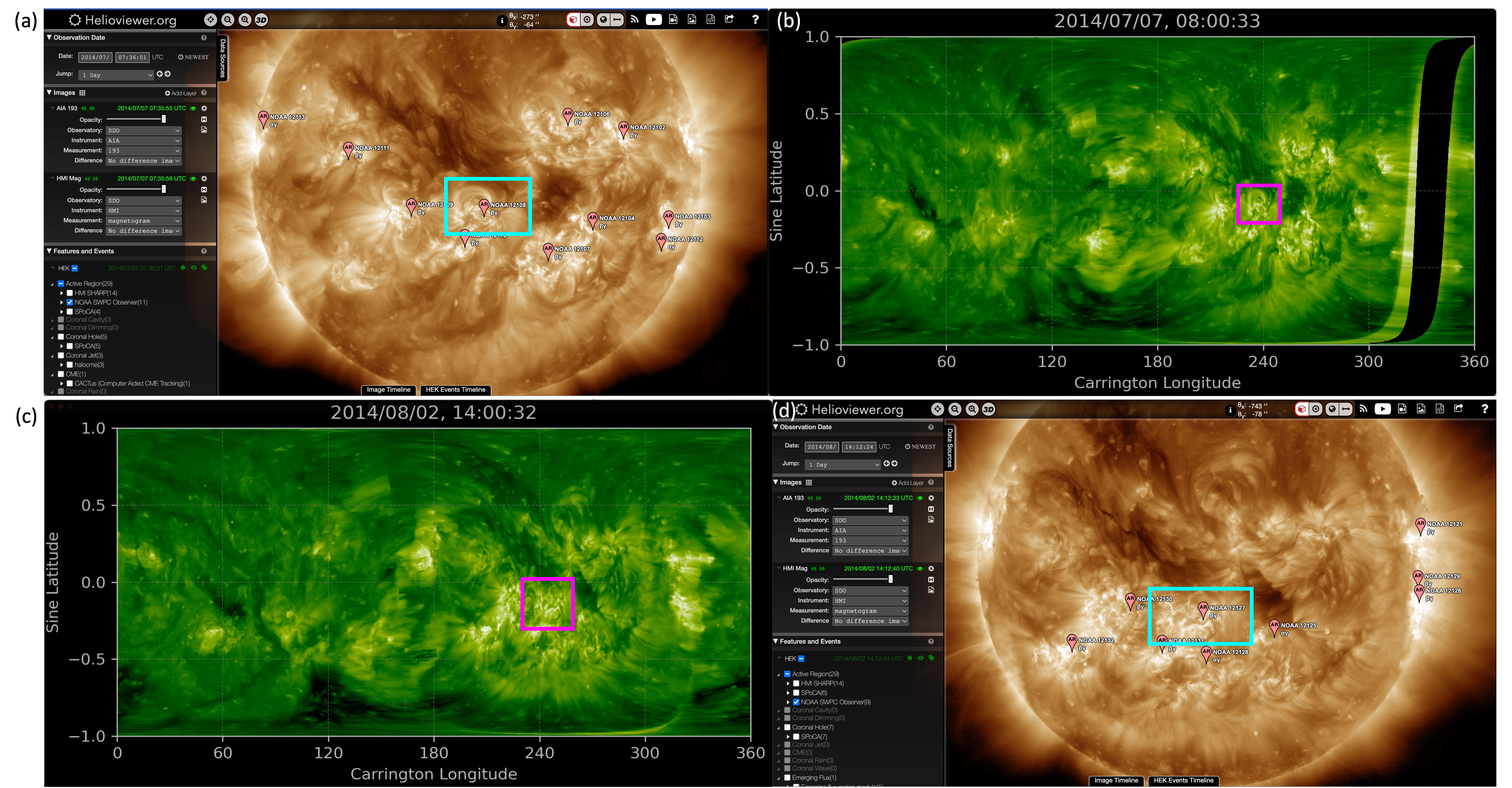}
    \caption{Example of the methodology used to identify LLARs. (a) A LLAR's initial NOAA AR designation, as displayed with Helioviewer with AR labels plotted from HEK. The LLAR is highlighted with a cyan box. (b) The same AR, highlighted with a magenta box, as confirmed through visual inspection in the CHMAP annual Carrington map movie. (c) One Carrington rotation later in the CHMAP movie, the LLAR still appears at the same location. (d) Identification of the new NOAA AR designation for the LLAR.}
    \label{fig:tracking_example}
\end{figure}

To evaluate this list of candidate pairs, we used the Carrington map annual movies of AIA and STEREO EUV imagery from the CHMAP Browser, which use the 195 Å data from STEREO and 193 Å data from AIA, with extensive instrument cross-calibration, at a cadence of 2 hours and a map size of 1600 x 640 pixels.\footnote{Available at \url{https://q.predsci.com/CHMAP-map-browser/}} We identified the AR to track by comparing the Carrington map movies to the data from the JHelioviewer and Solar Monitor projects, which had markers overplotted for the NOAA AR designations, whose locations were taken from the Heliophysics Event Knowledgebase (HEK).\footnote{\url{https://www.lmsal.com/hek/index.html}.} Once the correct AR had been identified, we visually tracked the active region to the next rotation and checked for a subsequent NOAA AR designation at the new location. For the years without full-Sun coverage, we supplemented the AR tracking and development described here with Stanford's far-side helioseismic synchronic maps, available from JSOC. An example case, for the LLAR pair NOAA AR 12108-12127, is shown in Figure \ref{fig:tracking_example}.

The definitive classifications have been made available via Zenodo at \url{https://doi.org/10.5281/zenodo.17298473}. All statistics and graphs shown in the following section have been calculated using this set of the LLAR identifications.

\subsection{AR Candidate Identification}\label{sec:ids}

% \subsubsection{Definitive Classifications}\label{sec:def_class}

To produce the list of confirmed LLARs that is now on Zenodo, we used synchronic EUV maps, full-disk EUV imagery, and far-side helioseismic data. Together these allowed full-Sun tracking, regardless of the availability of STEREO data. It is well-established that flux is more likely to emerge into an area that already had an AR \citep{Harvey1993,schrijver2003}, and in fact we eliminated approximately 100 LLAR candidates solely because they had clear major secondary emergence phases. Without full-Sun magnetograms, it is very difficult to fully eliminate LLARs that had new large-scale flux emergence during far-side transit; there are some methods, such as \cite{UgarteUrra2015} and \cite{Chen2022}, that can estimate the magnetic field using 304 Å data as a proxy. However, when the STEREO spacecraft were not at the optimal spacing during their orbits, and particularly after the loss of STEREO-B, there are still gaps which cannot be filled.. Therefore, if a LLAR candidate reappeared on the Earth-facing side with an obviously stronger or more complex magnetic configuration than it had on the previous rotation or without significant field dispersion, it was discounted for this reason. Since ARs can have a range in their decay rates, this was a highly restrictive condition. However, the ultimate goal of this dataset is to provide LLARs which can be studied over their full life cycle. Including ARs with large-scale flux emergence would skew studies on AR decay, so we erred on the side of caution. For this reason, we present the final set of LLARs as a representative, but likely not exhaustive, list.

% \subsubsection{Participatory Science Project}\label{sec:citsci}

\section{Results and Statistics}\label{sec:results}
\begin{figure}[htbp]
    \centering
    \includegraphics[width=0.55\linewidth]{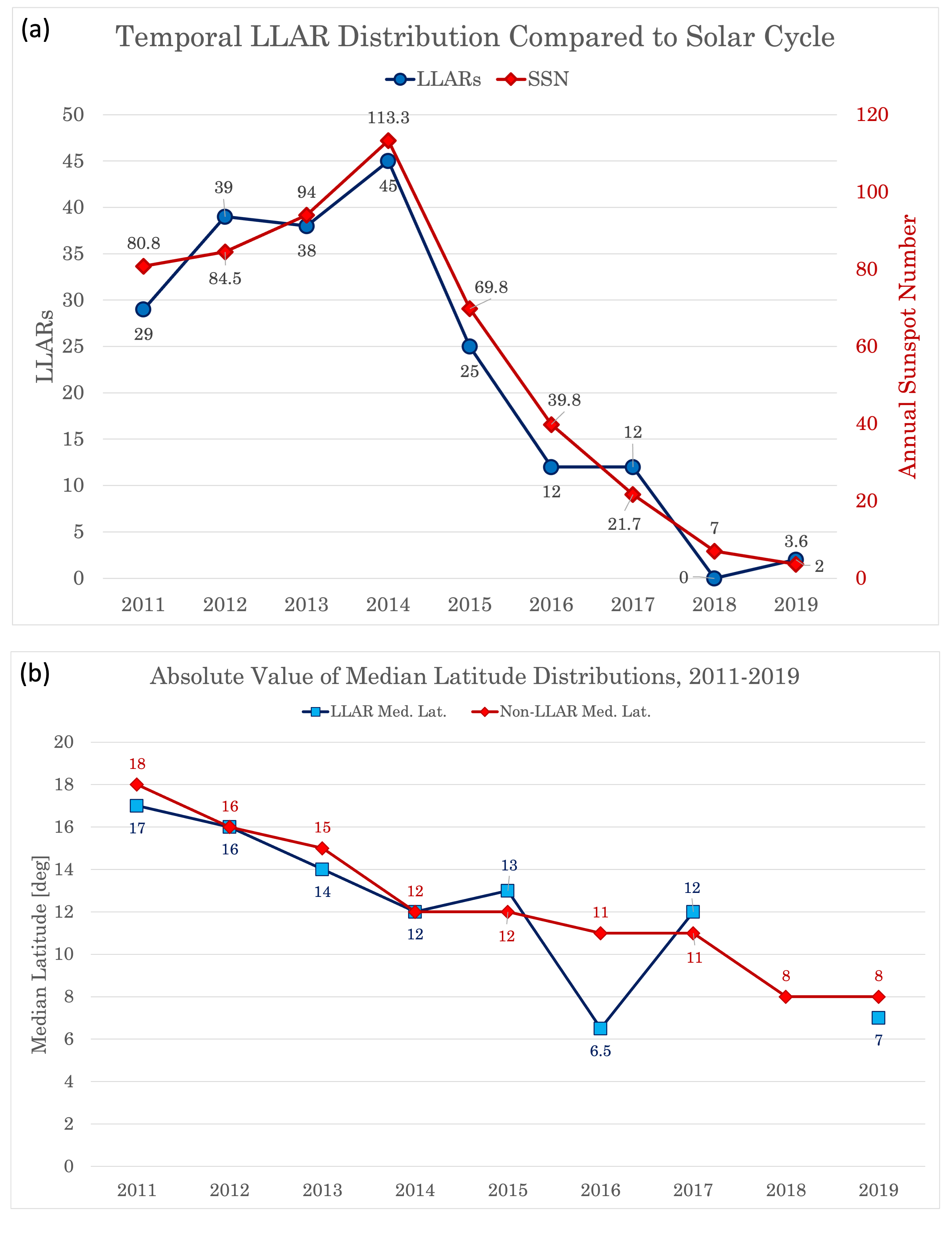}
    \caption{(a) Plot of the annual sunspot number and the number of LLARs detected in each year from 2011-2019. (b) Median latitudinal distribution of LLARs and non-LLARs in each year over the same period.}
    \label{fig:llars_sc}
\end{figure}

% go back and check this before submission, after spot-checking
Overall, in the course of this study, out of an initial input of all 1611 unique NOAA AR designations from 2011-2019 (collected from the HEK database using Fido in {\tt SunPy}), there were 101 LLARs identified, covering 214 individual NOAA AR numbers. From this, it is clear that the majority received two NOAA AR numbers on two Carrington rotations before either significant decay rendered the original AR undetectable in continuum emission, or new emergence discounted it from the list (as was the case for roughly 100 candidate pairs; these account for less than 10\% of the total non-LLAR pool, meaning the vast majority of non-LLARs were of shorter duration than their LLAR counterparts). In the following statistics, no ARs have been systematically removed from the calculations; some quantities were not available for all LLARs or non-LLARs (particularly the SHARP-derived quantities), but all available data was used in every calculation. 7 LLARs received 3 NOAA AR numbers, and 1 received 4 NOAA AR numbers. No ARs received more than 4 NOAA designations. The distribution of the LLARs with respect to the solar cycle, as well as their median latitudinal distribution compared to non-LLARs for each year of the study, are shown in Fig. \ref{fig:llars_sc}.

From this plot, it is clear that the distribution of LLARs closely mirrors the phase of the solar cycle; indeed, the Pearson coefficient for this comparison is 0.982. This implies that LLARs are subject to the same drivers for emergence and dissipation as other ARs (which is further bolstered by their very similar distribution in latitude throughout the solar cycle), and does not explain their longer lifespans. We investigated whether the LLAR's longevity could be explained by these ARs having an anti-Hale orientation, as this could suppress the well-known ``rush to the poles'' effect and potentially result in longer-lived flux concentrations. The distinction depends upon whether the AR follows Hale's Law \citep{Hale1925}, which simply states that the leading magnetic polarities dominant in the northern and southern solar hemispheres are oppositely-oriented during any given solar cycle (i.e., if negative polarity is leading in the northern hemisphere, positive polarity will lead in the southern hemisphere). ARs which follow this rule are designated as Hale, and those which oppose it -- often about 10\% or less of the ARs in a given cycle -- are termed anti-Hale. About 5\% of the LLARs were anti-Hale, which is in keeping with the range of 3-10\% found in the literature \cite{Richardson1948,Stenflo2012,Li2018,Zhukova2020}. Therefore, this is unlikely to be a major factor for the extended lifetime of LLARs.

\begin{figure}[htbp]
    \centering
    \includegraphics[width=0.6\linewidth]{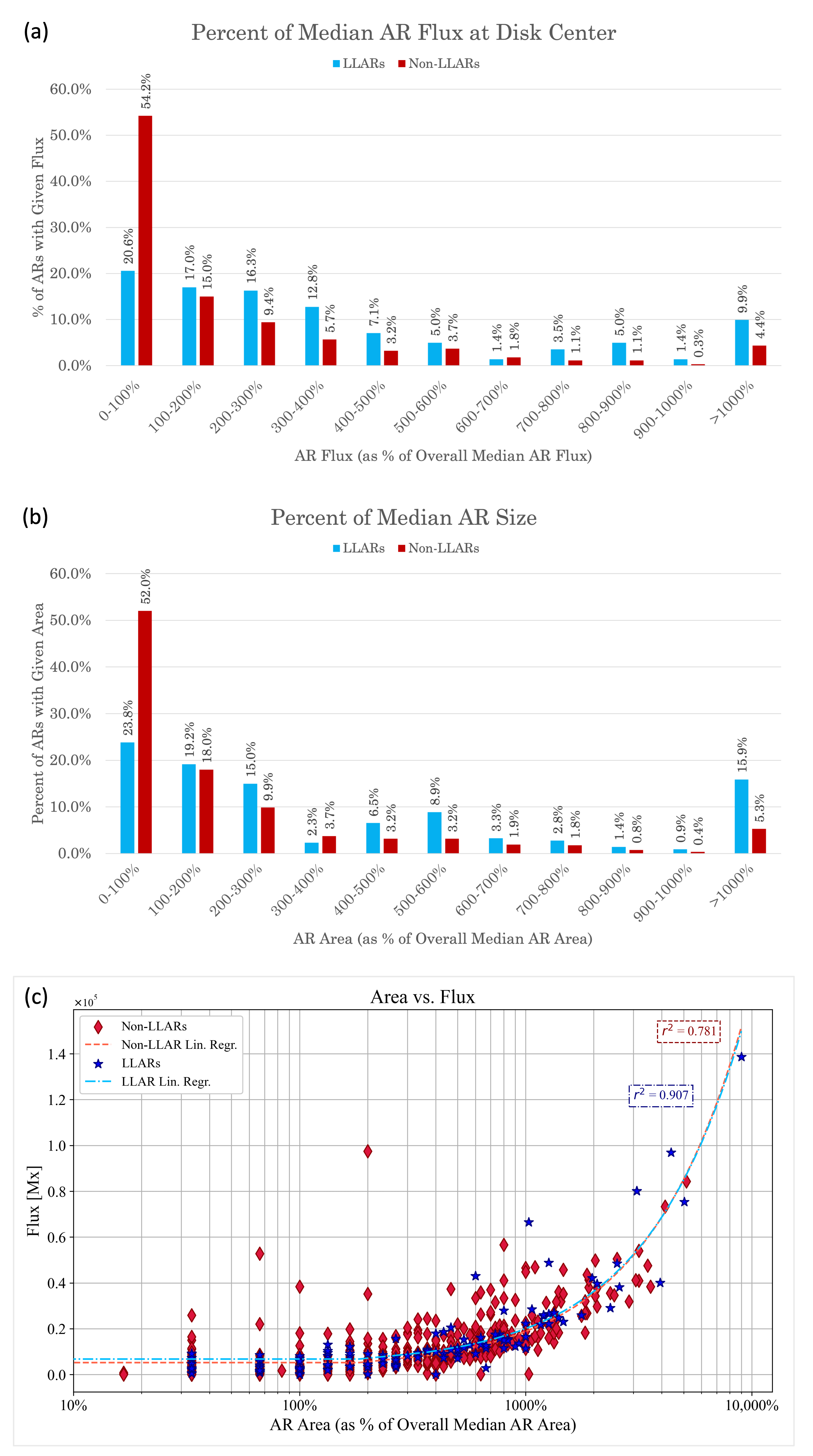}
    \caption{(a) Histogram comparing LLAR and non-LLAR magnetic flux by year and percent of the median AR flux (calculated over all ARs), in the years 2011-2019. (b) Analogous plot, but for area as a percent of the overall median AR size (calculated over all ARs). (c) Scatter plot showing the relationship between AR area and total unsigned flux for both LLARs and non-LLARs, and linear fits to both categories (the x axis is log-scaled). The r-squared value for the LLAR and non-LLAR fits are about 0.91 and 0.78, respectively. Note: the number of overlapping points in the 10-100\% range is very high; this plot is not in contradiction to the distributions seen in (b).}
    \label{fig:area_comps}
\end{figure}

We also investigated the relative characteristics of LLARs compared to non-LLARs, the latter of which were defined by using the entire list of labeled NOAA ARs from the same period which were not part of the LLAR list. The area and magnetic flux comparisons are summarized in Fig. \ref{fig:area_comps}. For all of the following calculations, the ``overall'' medians refer to those which were calculated using the entire dataset of ARs, regardless of whether they were classified as a LLAR or non-LLAR. The flux calculations were made using the total unsigned flux from SHARPs, taken at the closest available observation to disk center for each AR. LLARs tend to have significantly more concentrated magnetic field than their shorter-lived counterparts; over half (53.6\%) of LLARs have unsigned fluxes in the range of 1-5x the median AR flux, while 54.3\% of non-LLARs occupy the less-than 1x bin. The areas of LLARs also tend to be much greater than the majority of non-LLARs, with only 32\% of LLARs lying below the median AR size, compared to 63\% of non-LLARs. By contrast, 14\% of LLARs are 10 times the median AR size, only but 5\% of non-LLARs are. This is likely due to the fact that smaller ARs decay faster than large ARs, as previously found by \cite{Plotnikov2023}.

The relationship between area and total unsigned magnetic flux (as derived by SHARP data, when available, taken near disk center) is directly linear, and very close between LLARs and non-LLARs. Both have high $r^2$ values, and very low residual standard errors ($S_{LLAR}\approx64$ and $S_{Non-LLAR}\approx44$ Mx, respectively). LLARs have a slightly lower slope, which is unsurprising given the much greater range of areas present in this population.

\begin{figure}[htbp]
    \centering
    \includegraphics[width=0.75\linewidth]{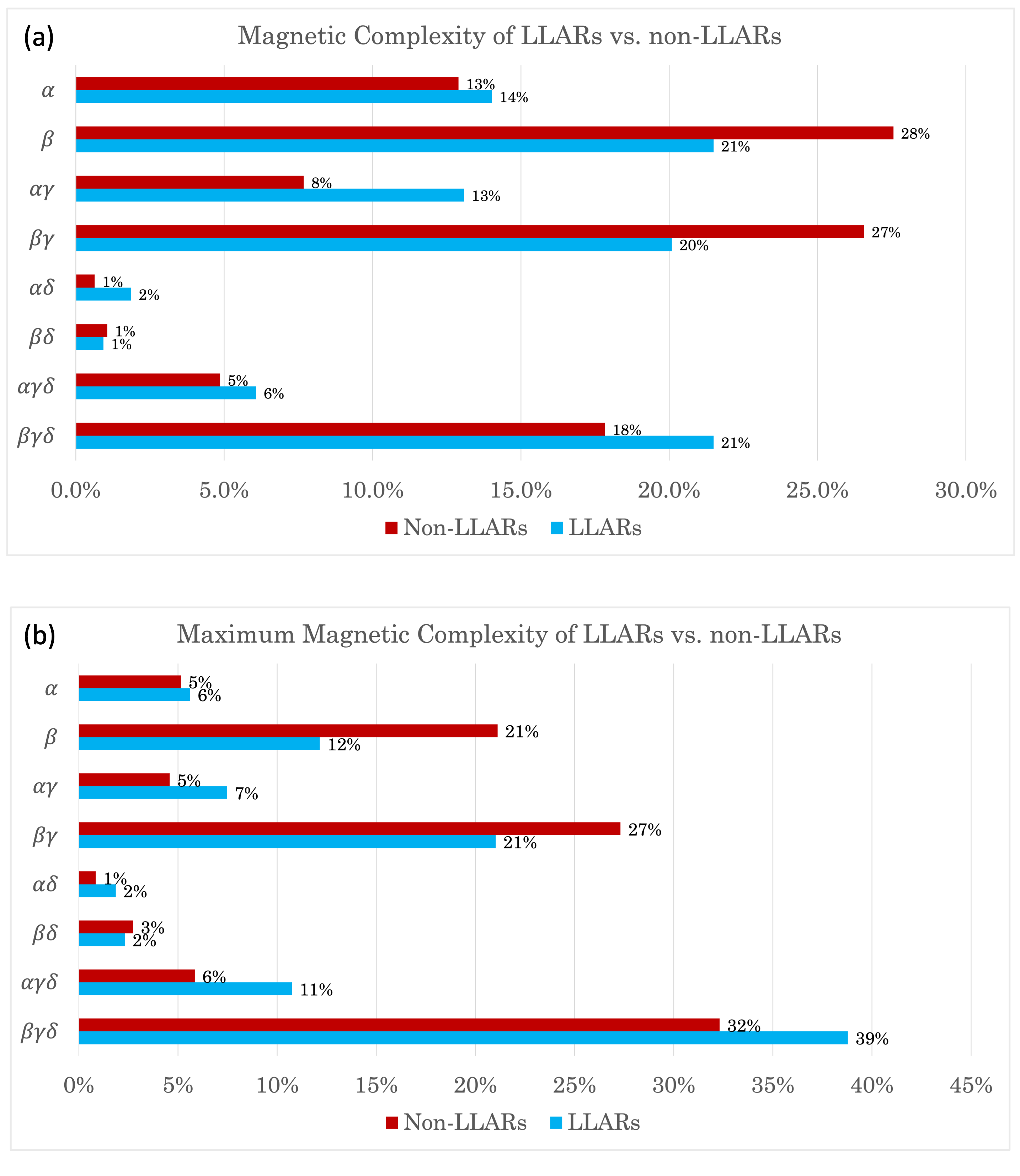}
    \caption{(a) Distribution of magnetic complexity, as measured by Mt. Wilson class at disk-center passage, across LLARs and non-LLARs. (b) The same plot, but for the maximum Mt. Wilson class recorded during front-side passage.}
    \label{fig:mag_complex}
\end{figure}

Critically, LLARs are not significantly more magnetically complex than non-LLARs, as seen in Fig. \ref{fig:mag_complex}. Overall there is no coherent trend that distinguishes the two populations among the Mt. Wilson classification scheme. Magnetic configuration can, of course, evolve drastically during an AR's lifetime, and single identifications (which, like our other measurements, were taken from the designation at the closest available time to disk-center) are unlikely to accurately capture the range of classifications each AR may have had during its disk passage. Therefore, we also include the distributions for the maximum Mt. Wilson designation achieved by each AR during its entire front-side passage. Neither shows a strong correlation between complexity and LLAR designation; we discuss this peculiar finding in greater detail in Section \ref{sec:disc}.

Perhaps the most dramatic result from this assessment of these LLARs is that they are disproportionately more flare-active than non-LLARs. Table \ref{tab:RAR_flare_stats} shows the average number of flares for all flares (taken from the Hinode XRT flare catalog, which covers A through X class flares), and then for C, M, and X flares separately. The grouped ratios count the flares by the 101 unique ARs classified as LLARs, while the individual ratio count by the 214 individual NOAA designations the LLARs received. In brief, LLARs are nearly 4 times as likely to host a C-class flare, and that factor increases to over 5 for M-class and over 6 for X-class flares.
\begin{table}[]
\centering
\begin{tabular}{|ll|}
\hline
\multicolumn{2}{|c|}{\textbf{Average Number of Flares}}                    \\ \hline
\multicolumn{1}{|l|}{\textit{Grouped LLARs}} & 24.59                       \\ \hline
\multicolumn{1}{|l|}{\textit{Indiv. LLARs}}  & 11.37                       \\ \hline
\multicolumn{1}{|l|}{\textit{Non-LLARs}}     & 6.50                        \\ \hline
\multicolumn{1}{|l|}{\textit{Grouped Ratio}} & {\color[HTML]{FF0000} 3.78} \\ \hline
\multicolumn{1}{|l|}{\textit{Indiv. Ratio}}  & {\color[HTML]{FF0000} 1.75} \\ \hline
\multicolumn{2}{|l|}{}                                                     \\ \hline
\multicolumn{2}{|c|}{\textbf{Average Number of C-class Flares}}            \\ \hline
\multicolumn{1}{|l|}{\textit{Grouped LLARs}} & 14.78                       \\ \hline
\multicolumn{1}{|l|}{\textit{Indiv. LLARs}}  & 6.80                        \\ \hline
\multicolumn{1}{|l|}{\textit{Non-LLARs}}     & 3.80                        \\ \hline
\multicolumn{1}{|l|}{\textit{Grouped Ratio}} & {\color[HTML]{FF0000} 3.89} \\ \hline
\multicolumn{1}{|l|}{\textit{Indiv. Ratio}}  & {\color[HTML]{FF0000} 1.79} \\ \hline
\multicolumn{2}{|l|}{}                                                     \\ \hline
\multicolumn{2}{|c|}{\textbf{Average Number of M-class Flares}}            \\ \hline
\multicolumn{1}{|l|}{\textit{Grouped LLARs}} & 1.96                        \\ \hline
\multicolumn{1}{|l|}{\textit{Indiv. LLARs}}  & 0.88                        \\ \hline
\multicolumn{1}{|l|}{\textit{Non-LLARs}}     & 0.36                        \\ \hline
\multicolumn{1}{|l|}{\textit{Grouped Ratio}} & {\color[HTML]{FF0000} 5.47} \\ \hline
\multicolumn{1}{|l|}{\textit{Indiv. Ratio}}  & {\color[HTML]{FF0000} 2.47} \\ \hline
\multicolumn{2}{|l|}{}                                                     \\ \hline
\multicolumn{2}{|c|}{\textbf{Average Number of X-class Flares}}            \\ \hline
\multicolumn{1}{|l|}{\textit{Grouped LLARs}} & 0.16                        \\ \hline
\multicolumn{1}{|l|}{\textit{Indiv. LLARs}}  & 0.07                        \\ \hline
\multicolumn{1}{|l|}{\textit{Non-LLARs}}     & 0.02                        \\ \hline
\multicolumn{1}{|l|}{\textit{Grouped Ratio}} & {\color[HTML]{FF0000} 6.73} \\ \hline
\multicolumn{1}{|l|}{\textit{Indiv. Ratio}}  & {\color[HTML]{FF0000} 2.93} \\ \hline\end{tabular}
\caption{Average flare counts for LLARs as individual ARs, LLARs grouped by identity (the sum of the flares for all of the NOAA designations a LLAR received), and non-LLARs. The ratios at the bottom of each category in red are the ratios of the LLAR categories to the non-LLAR category.}
\label{tab:RAR_flare_stats}
\end{table}

In keeping with these results, LLARs show significantly higher median values then other ARs for many of the SHARP summary parameters. Several of these parameters are summarized in Table \ref{tab:sharp_stats}. These quantities are well known to be correlated with overall flare productivity \citep{Bobra2014,Chen2019}, so these findings serve primarily as a verification of the LLARs' characteristics found in this study.

\begin{table}[]
\begin{tabular}{|c|c|c|c|}
\hline
\textit{}                                     & LLAR         & Non-LLAR     & LLAR/Non-LLAR \\ \hline
Absolute Net Current Helicity ($G^2\, m^{-1}$)  & 71.1  & 17.8  & 3.99          \\ \hline
Mean Current Helicity ($G^2\, m^{-1}$)          & -2.02 $\cdot\,10^{-3}$ & -3.57 $\cdot\,10^{-4}$  & 5.67          \\ \hline
Mean Phot. Magnetic Energy Density ($erg\, cm^{-3}$) & 6930 & 3080 & 2.25 \\ \hline
Mean Shear Angle ($deg$)                        & 39.3  & 31.8  & 1.23          \\ \hline
Mean Twist ($M\, m^{-1}$)                     & -4.50 $\cdot\,10^{-3}$ & -1.42 $\cdot\,10^{-3}$ & 3.16          \\ \hline
Mean Vertical Current Density ($mA\, m^{-2}$) & 0.326  & 0.347  & 0.94          \\ \hline
Unsigned Current Helicity ($G^2\, m^{-1}$) & 679  & 214  & 3.17          \\ \hline
Unsigned Vertical Current ($A$)                 & 1.62$\cdot\,10^{13}$  & 4.87$\cdot\,10^{12}$  & 3.32          \\ \hline
\end{tabular}
\caption{Table of median SHARP parameter values for LLAR and non-LLAR ARs, as well as their ratios, for the NOAA designations which had corresponding SHARP data.}\label{tab:sharp_stats}
\end{table}

\section{Discussion and Conclusions}\label{sec:disc}

In this paper we present a dataset of ARs that were identified with NOAA AR designations at least twice (and up to four times) without undergoing significant additional magnetic flux emergence during their decay phases. Our goal is to facilitate a more statistical approach to studying AR evolution and decay within the community. The full list is available on Zenodo at \url{https://doi.org/10.5281/zenodo.17298473}.

Two quantities listed in Table \ref{tab:sharp_stats} have a weaker LLAR/non-LLAR ratio than the rest: mean shear angle and mean vertical current density. The values from which the table was made are median values for all available LLARs and non-LLARs in the SHARP database \textit{at the time the AR was nearest disk center}. This last part is relevant because, unlike most of the other values in the table, component values and the shear angle are likely to evolve on relatively short time frames. Furthermore, both are known to evolve rapidly before, during, and after flares (see \cite{Kniezewski2025} for a deeper discussion on the pre-flare evolution of these values).

Another intriguing facet of LLARs is their temporal distribution throughout the solar cycle. They generally follow the annualized sunspot number as closely as non-LLARs, which is slightly counterintuitive. All LLARs which persisted but which had significant flux emergence during subsequent rotations were removed from the dataset, which means that LLARs are even more prevalent closer to solar maximum than is reflected in the final results. One would naïvely expect, then, that the resultant distribution might be more bimodal, showing the most LLARs during the increasing and decreasing phases, but not actually near solar maximum itself. There is a slight increase in the number of LLARs during these phases; however, given the low numbers of LLARs overall, more solar cycles of data would be required in order to confirm this correlation. 

Alternatively, it is possible that the crowding of many localized, concentrated fields during solar maximum biases towards a longer decay time for a single AR, while ARs in relative isolation during quieter periods decay more readily even if they are relatively large and high-flux. We will continue to monitor LLARs in the current solar cycle 25 to extend the analysis. We have also undertaken a study of 5 LLARs which were observed for 3 rotations each, in order to assess changes to the heating cycles within each LLAR as they decayed. That work will be presented in an upcoming publication.

We have reserved the most important discussion for last. In characterizing some basic qualities of the LLARs during the preliminary analysis of the dataset, we found that they tend to have a significantly broader distribution of areas, but that they are not much more magnetically complex than other ARs. However, LLARs are statistically 3-6 times more likely to produce flares of every category, and they have correspondingly greater median values for most of the SHARP derived quantities for flare activity correspondence. While it is not a new finding that AR size and flux are closely related to overall flare productivity, the actual level of flare productivity we found in LLARs was remarkable. Furthermore, we find it highly intriguing that LLARs do not conform to the well-established, positively correlated relationship between magnetic complexity and solar flare activity seen in the majority of active regions. After considering total area, location, magnetic flux, magnetic complexity, and an array of flare-correlative measures, there is simply no clear explanation for LLARs to consistently break this rule of thumb. Since all LLARs account for less than 7\% of all unique ARs (101 out of 1521), there are too few of them to disturb the general correlation, and we do not suggest that the correlation is flawed. However, neither does this explain why LLARs are so different. The Mt. Wilson categorization scheme is known to over-classify regions into complex designations, but this skews results in the opposite direction, strengthening the usual complexity/flare relationship rather than weakening it.

The increased flux across the LLARs may provide a hint to this puzzle; it is possible that LLARs are formed from stronger flux regions rooted more deeply in the sub-surface regions, lending both to their longevity and their eruptive activity levels. It might be anticipated that their rotation time would therefore differ from the expected rotation time for their respective formation latitudes. Accordingly, we calculated the difference between the expected time of rotation for each LLAR and the actual time of rotation, and found that LLARs rotate slightly faster than the predicted rotation time, with a median time of 5 hours faster. As this is less than 1\% difference from the predicted period, however, it is unlikely to be physically meaningful. This conundrum makes a strong argument for further study of these ARs, both for flare research in general and to be able to identify them early for long-range flare predictions in space weather operations. LLARs constitute a small but diverse and operationally important category of active coronal phenomena.

\section{Acknowledgments}

EIM was supported for this work by NASA grants 80NSSC23K0159 and 80NSSC25K7681. The authors would like to thank David Fritz for his contribution to the project in developing the original version of the LLAR candidate identifying script.

\bibliography{rar_intro_bib}{}
\bibliographystyle{aasjournalv7}

\appendix

\begin{figure}[htbp]
    \centering
    \includegraphics[width=0.8\textwidth]{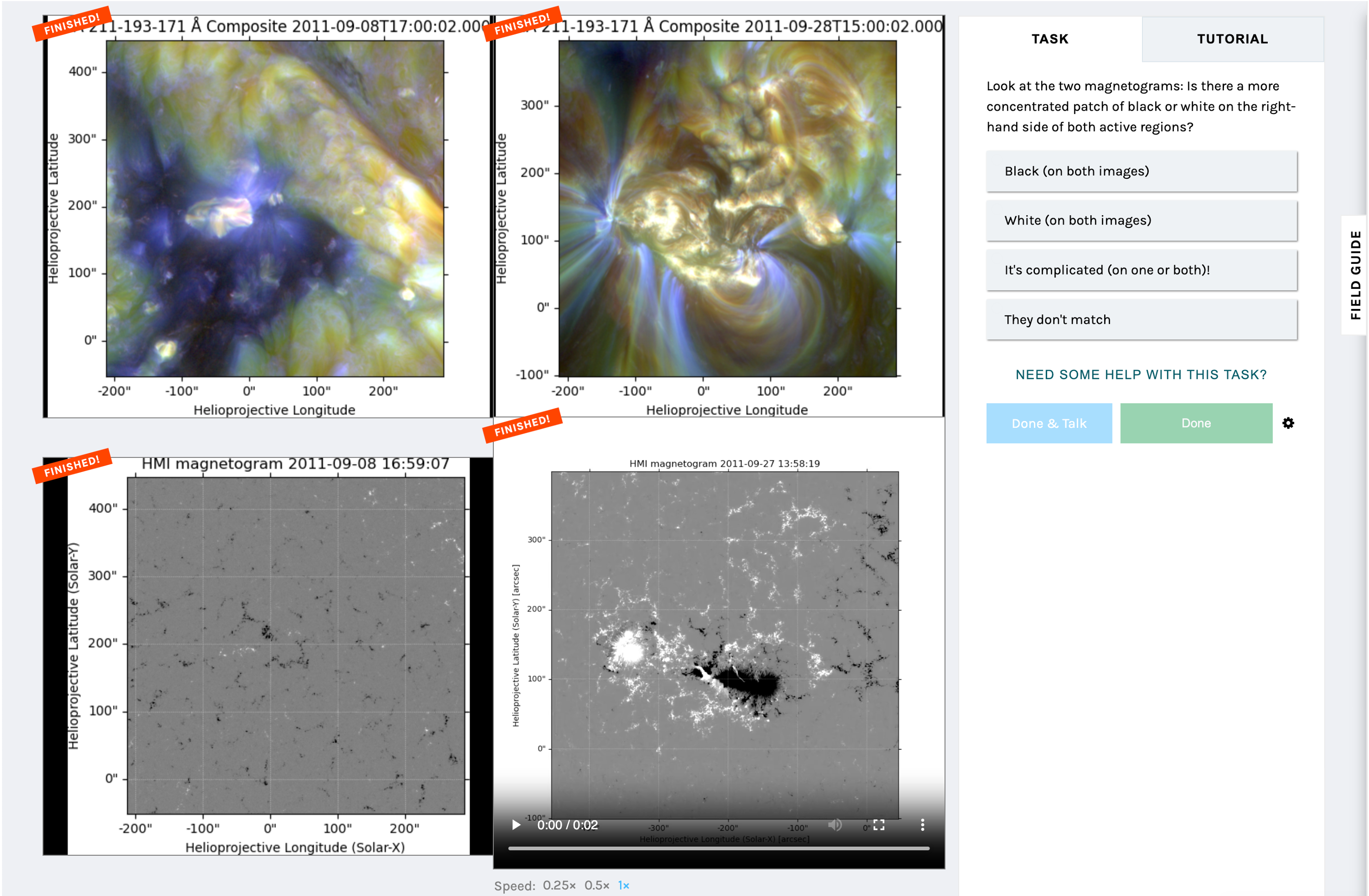}
    \caption{Sample image of the layout of the Solar Active Region Spotters project on Zooniverse. The top row shows the SDO AIA composite images for the two candidates (this case is not a match), while the bottom left shows a single HMI magnetogram of the original AR and the bottom right shows a short HMI movie of the subsequent AR's evolution. The red flags in the top left of each figure denote that the subject pair shown has already garnered enough classifications to be retired.}
    \label{fig:zoon}
\end{figure}

As part of this project, we created and managed a participatory science project called Solar Active Region Spotters, which was hosted on the widely used citizen science platform Zooniverse. The purpose of the participatory science aspect was twofold: to gather details about the ARs which were not easily derived through image processing (e.g., prominence identification and evolution), but that might shed light on LLARs; and to evaluate the efficacy of crowd-sourced identifications for more complex data sets, when compared with expert identifications of the same pool. We downloaded 500x500 arcsec cutouts of each AR in the SDO AIA channels 171, 193, and 211 Å, and created composite images from them, along with single frames of SDO HMI magnetogram cutouts for the intial ARs and brief HMI cutout movies (2 days at half-hour cadence) showing the subsequent rotation's candidates. These were used as the dataset, an example view of the workflow for which is shown in Fig. \ref{fig:zoon}. Over 1300 volunteers classified the 576 candidate pairs 15 times each; this was the threshold set to collect sufficient statistics in order to retire each case. Volunteers were asked 3 questions:
\begin{itemize}
    \item Look at the two magnetograms: Is there a more concentrated patch of black or white on the right-hand side of both active regions?
    \item Look at the two multicolor EUV images: Can you see one or more prominences in the active region? 
    \item Do you think the active region in the movie on the right could be an older version of the one in the image on the left?
\end{itemize}
The first question's aim is to help determine whether the ARs in question were Hale or anti-Hale in nature. Of course, if the two ARs' Hale categorizations differ, then they cannot be a single LLAR, so this question also aided in the identification process. The goal of the second question is to identify internal structure that may help in verifying the pair as a LLAR. Prominences present in the later image could well have developed during the intervening rotation, though it is less likely to be useful if a prominence is present in the first AR and not in the second (it could be a sign that they were different ARs, or it could simply mean that the prominence material erupted at some point before the second image). The presence of a distinctive prominence shape and orientation in both images can be helpful, though by itself insufficient, evidence for confirming a LLAR. The third question is the main result, which was intended to tap into volunteers' ``gut feeling'' about the pair, based on the initial training material they were given on AR evolution.

Upon completing the definitive classifications, discussed in Section \ref{sec:ids}, we evaluated the results of the participatory science project. The accuracy for the overall identification question (defined as $\frac{true\, positive + true\, negative}{total \,classifications}$) was around 64\%, but the precision ($\frac{true\,positive}{true\,positive+false\,positive}$) was 0.07. For the question of which polarity dominated on the right-hand side of each LLAR, the volunteer responses scored at 53\% accuracy and 0.91 precision. Additionally, the responses to the question on the presence or absence of prominences came in at 58\% accuracy and 0.78 precision. Unfortunately, such low accuracy meant that the results were not able to be utilized in synthesizing the final LLAR list. However, these values are unsurprising, as the tasks were of relatively high difficulty compared to most other participatory science investigations. Projects such as Solar Jet Hunters \citep{Musset2024} are effective because they focus on simple tasks that are quickly learned, such as drawing boxes around distinct, bright features. The volunteers for this project generally had no prior background in solar observations and were required to learn about magnetograms, EUV observations, coronal loops, filaments, and AR evolution, all prior to beginning the classification task. The volunteers nevertheless showed a high level of engagement and their discussions in the forum showed that they had gained a significant understanding of the corona in general, so for outreach and educational purposes, the Zooniverse project was successful.

\end{document}